\documentclass[twocolumn,linenumbers]{aastex631}
\usepackage{natbib}
\usepackage{amsmath}
\usepackage{appendix}
\makeatletter
\def\@fnsymbol#1{\@arabic{#1}}
\makeatother

\begin{document}
\nolinenumbers

\author[0000-0002-8466-5469]{Collin Cherubim}
\affiliation{Department of Earth and Planetary Sciences, Harvard University, 20 Oxford St., Cambridge, MA 02138, USA}
\affiliation{Center for Astrophysics \textbar \ Harvard \& Smithsonian, 60 Garden St., Cambridge, MA 02138, USA}

\author{Robin Wordsworth}

\affiliation{School of Engineering and Applied Sciences, Harvard University, 20 Oxford St., Cambridge, MA 02138, USA}

\affiliation{Department of Earth and Planetary Sciences, Harvard University, 20 Oxford St., Cambridge, MA 02138, USA}

\author{Renyu Hu}
\affiliation{Jet Propulsion Laboratory, California Institute of Technology, Pasadena, CA 91109, USA}
\affiliation{Division of Geological and Planetary Sciences, California Institute of Technology, Pasadena, CA 91125, USA}

\author{Evgenya Shkolnik}
\affiliation{School of Earth and Space Exploration, Arizona State University, Tempe, AZ 85281, USA}
\affiliation{Center for Astrophysics \textbar \ Harvard \& Smithsonian, 60 Garden St., Cambridge, MA 02138, USA}

\correspondingauthor{Collin Cherubim}
\email{ccherubim@g.harvard.edu}

\title{Strong fractionation of deuterium and helium in sub-Neptune atmospheres along the radius valley}

\begin{abstract}
We simulate atmospheric fractionation in escaping planetary atmospheres using {\tt\string IsoFATE}, a new open-source\footnote{IsoFATE source code: \url{https://github.com/cjcollin37/IsoFATE}} numerical model. We expand the parameter space studied previously to planets with tenuous atmospheres that exhibit the greatest helium and deuterium enhancement. We simulate the effects of EUV-driven photoevaporation and core-powered mass loss on deuterium-hydrogen and helium-hydrogen fractionation of sub-Neptune atmospheres around G, K, and M stars. 
Our simulations predict prominent populations of deuterium- and helium-enhanced planets along the upper edge of the radius valley with mean equilibrium temperatures of $\approx$ 370 K and as low as 150 K across stellar types. We find that fractionation is mechanism-dependent, so constraining He/H and D/H abundances in sub-Neptune atmospheres offers a unique strategy to investigate the origin of the radius valley around low-mass stars. Fractionation is also strongly dependent on retained atmospheric mass, offering a proxy for planetary surface pressure as well as a way to distinguish between desiccated enveloped terrestrials and water worlds. Deuterium-enhanced planets tend to be helium-dominated and CH$_4$-depleted, providing a promising strategy to observe HDO in the 3.7 $\mu$m window. We present a list of promising targets for observational follow-up.
\end{abstract}

\section{Introduction} \label{sec:intro}

The radius distribution of small exoplanets is bimodal \citep{Fulton_2017, Fulton_2018, Mayo_2018, Eylen_2018, Berger_2020, Cloutier_2020}. Theoretical and observational evidence suggests that the radius valley is sculpted in large part by thermally-driven atmospheric escape of primordial atmospheres \citep{Owen_2013, Lopez_2014, Jin_2014, Owen_2017, Jin_2018, Ginzburg_2018, Gupta_2019, Gupta_2020, Eylen_2021, Rogers_2023, Berger_2023}, while accretion of water-rich ices \citep{Luque_2022} and/or formation timescale \citep{Lee_2014, Lopez_2018, Lee_2021, Cherubim_2023} may also play significant roles, especially for planets around low-mass stars \citep{Cloutier_2020}. These processes produce two broad populations of planets with distinct mass-radius profiles: terrestrials and enveloped terrestrials, the former being Earth-like rocky planets and the latter likely possessing an extended H/He envelope and/or water-dominated ices.

Extreme ultraviolet (EUV)-driven photoevaporation and core-powered mass loss are two important proposed mechanisms for driving hydrodynamic escape of planetary atmospheres \citep{Sekiya_1980, Watson_1981, Ginzburg_2018}. Photoevaporation results from atmospheric absorption of high-energy radiation originating from the host star and subsequent hydrodynamic outflow in the form of a Parker wind, the bulk of which occurs in the first several hundred Myr post-formation for most stellar types \citep{Parker_1965}. Core-powered mass loss is instead driven by a combination of host star bolometric luminosity and remnant planet formation energy and drives hydrodynamic outflow over Gyr timescales. Both mechanisms may have profound impacts on atmospheric composition in part due to mass fractionation of atmospheric constituents, including isotopes \citep{Sekiya_1980b, Watson_1981, Hunten_1987, Zahnle_1990, Yung_2000}. 

Atmospheric fractionation is of particular interest for sub-Neptunes (i.e. $R_\mathrm{p} \lesssim 4.0 R_\oplus$), the most common planetary class in the galaxy. Model predictions suggest that sub-Neptunes may become enriched in helium through escape \citep{Hu_2015, Malsky_2020, Malsky_2023}. Transmission spectroscopy observations of escaping helium tails with supersolar H:He ratios on sub-Neptunes support these predictions \citep{Zhang_2023, Orell-Miquel_2023}. Deuterium-protium (D/H) fractionation is of great interest as a tracer of planet formation and atmospheric evolution, and is potential observable with JWST and high-dispersion ground-based instruments \citep{Lincowski_2019, Molliere_2019, Morley_2019, Kofman_2019, Gu_2023}. Constraining D/H can also help assess planetary chemistry and habitability, since hydrogen loss is linked to both water loss and planetary oxidation \citep{Genda_2008,Wordsworth_2018,wordsworth2022atmospheres}.

Here we explore the differences in deuterium and helium fractionation of sub-Neptune atmospheres between thermally-driven escape mechanisms -- EUV-driven photoevaporation and core-powered mass loss -- using our newly developed numerical model, {\tt\string IsoFATE}: Isotopic Fractionation of ATmospheric Escape. Building on previous calculations in \cite{Hu_2015}, \cite{Malsky_2020}, \cite{Malsky_2023}, and \cite{Gu_2023}, we expand the analysis across a wider parameter space, introduce core-powered mass loss effects, and simulate much higher levels of fractionation (D/H mole ratios exceeding Venusian D/H $\approx$ 1,000x Solar in helium-dominated atmospheres). We find that the two escape mechanisms predict helium and deuterium enhancement in distinct regions of planet mass-radius-instellation space, offering a unique means to identify which mechanism dominates the evolution of sub-Neptune atmospheres. We describe the model in Sections \ref{sec:atmescape} and \ref{sec:isotopic_fractionation} and present our results in Section \ref{sec:results}. We discuss potential applicability to observational surveys in Section \ref{sec:Observation}, and state our conclusions in Section \ref{sec:Conclusion}.

\section{Atmospheric Escape Model} \label{sec:atmescape}

{\tt\string IsoFATE} simulates mass fractionation via two atmospheric escape mechanisms: EUV-driven hydrodynamic escape and core-powered mass loss, described in the following sections. In both cases, planetary radius evolves temporally as a result of atmospheric loss and thermal contraction. The overall mass loss rate is governed by Equations \ref{eq:EUVflux}, \ref{eq:phi_RR}, \ref{eq:phi_L}, and \ref{eq:phi_B}, defined in the following sections. Radius evolution due to thermal contraction on the other hand is governed by thermal evolution models presented in \cite{Lopez_2014}. This model component is derived from hydrostatic structure models for rocky planets enveloped in Solar-composition H/He atmospheres, assuming one third of the mass is in an iron core and two thirds is in a silicate mantle, which we assume too.

\subsection{EUV-driven hydrodynamic escape}
\label{sec:EUV}

\begin{figure}
    \centering
    \includegraphics[width=1.1\hsize]{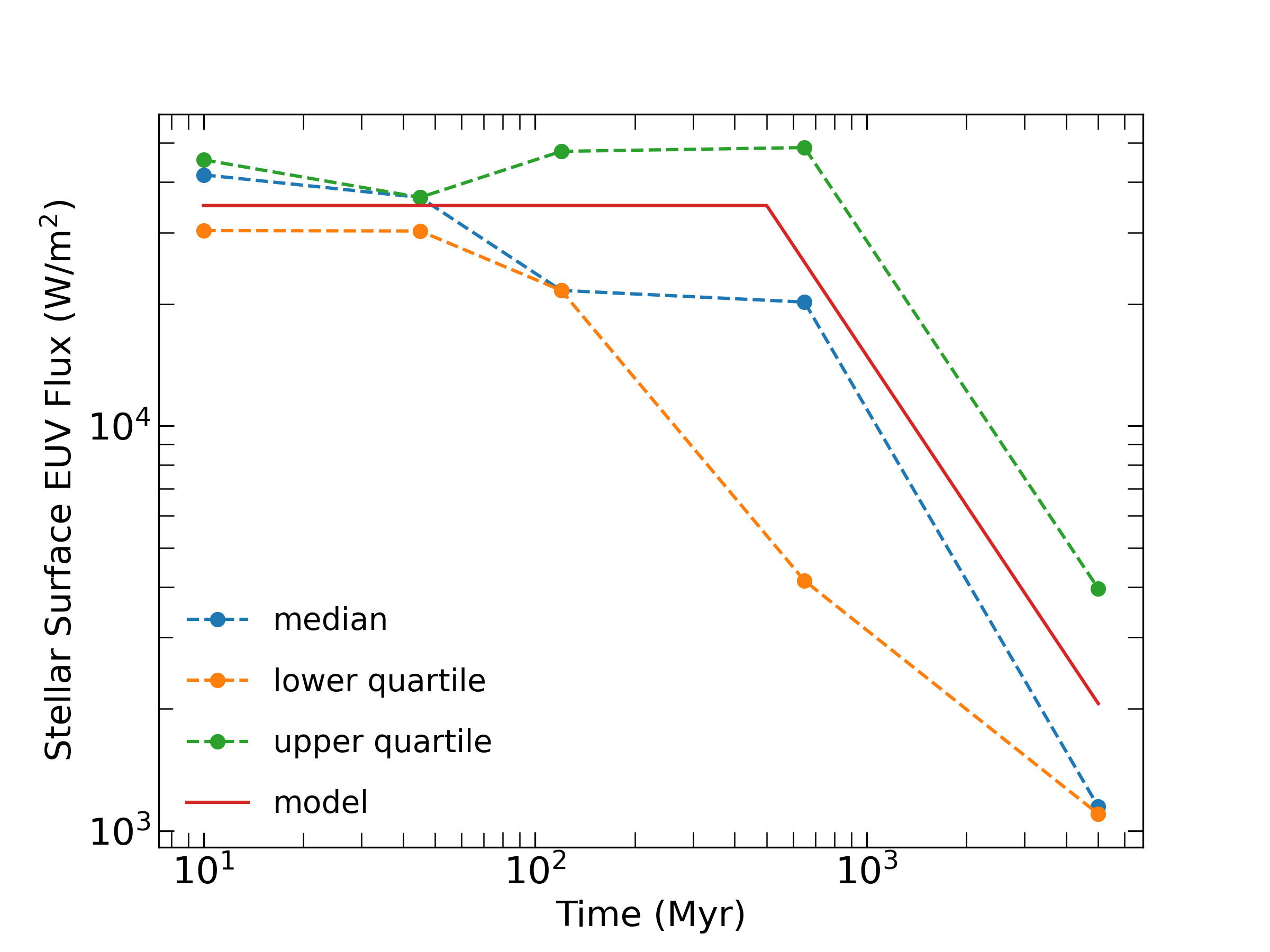}
    \caption{The solid red line shows the stellar surface EUV flux evolution model used in our photoevaporation simulations for planets around M stars (Section \ref{sec:EUV}, Equation \ref{eq:flux}). The dashed lines show the integrated flux density for 10 nm $<$ $\lambda$ $<$ 130 nm from the semi-empirical HAZMAT spectra \citep{Peacock_2020}. These spectra are computed for the lower quartile, median, and upper quartile EUV flux density samples of early M stars at five ages: 10 Myr, 45 Myr, 120 Myr, 650 Myr, and 5 Gyr. We chose $F_\mathrm{EUV, 0}$ = 3500 W m$^{-2}$ for our model, approximately equal to 0.1\% of the bolometric stellar surface flux.}
    \label{fig:EUV_model}
\end{figure}

Our stellar radiation model assumes EUV (10 nm $< \lambda <$ 130 nm) radiation from the host star is the primary driver of mass loss for the EUV-driven hydrodynamic escape mechanism. We ignore X-ray irradiation since approximately 80-95\% of the high energy flux is in the EUV \citep{Peacock_2020}. This also allows us to remain consistent with available semi-empirical spectra, as discussed below \citep{Loyd_2016, Fontenla_2016, Peacock_2019, Peacock_2019b, Peacock_2020, Richey-Yowell_2023}. We assume a H/He composition with solar metallicity (mean molecular mass = 2.35 $u$) in the bulk atmosphere and we assume that hydrogen in the escaping wind is atomic due to photodissociation of H$_2$ catalyzed by water vapor \citep{Liang_2003, Yelle_2004, Moses_2011, Hu_2012}.

To model the temporal evolution of the incident EUV flux, $F_\mathrm{EUV}$, for planets around M stars, we constructed a power law function as in \cite{Cherubim_2023}, \cite{Cloutier_2023}, and \cite{Cadieux_2023}. Our model is informed by semi-empirical spectra generated by the HAZMAT team for populations of M1 stars over a range of ages from 10 Myr to 5 Gyr \citep[Figure \ref{fig:EUV_model};][]{Peacock_2020}:

\begin{equation}
    F_\mathrm{EUV} = 
    \begin{cases}
    F_\mathrm{EUV, 0} & t < t_\mathrm{sat} \\
    F_\mathrm{EUV, 0} \left(\frac{t}{t_\mathrm{sat}} \right)^\beta & t > t_\mathrm{sat},
    \end{cases}
\label{eq:flux}
\end{equation}

\noindent where $t_\mathrm{sat}$ is the saturation time marking the transition from constant EUV flux to power law decay. Note that $F_\mathrm{EUV}$ is related to stellar EUV luminosity $L_\mathrm{EUV}$ by: $F_\mathrm{EUV} = L_\mathrm{EUV}/4 \pi a^2$, where $a$ is the orbital semi-major axis. We found $t_\mathrm{sat} = 500$ Myr to be consistent with the HAZMAT spectra and $\beta = -1.23$ to be consistent with the HAZMAT spectra as well as previously reported Solar data \citep{Ribas_2005}. We set $F_\mathrm{EUV, 0}$ equal to $0.1\%$ of the incident planetary flux, $F_\mathrm{p}$, in our model simulations to remain consistent with HAZMAT data. For planets around K dwarfs, we adopt $t_\mathrm{sat} = 200$ Myr, consistent with high energy spectra for stars of various ages in the HAZMAT survey \cite{Richey-Yowell_2023}. For planets around G dwarfs, we adopt $t_\mathrm{sat} = 100$ Myr \citep{Ribas_2005, Garces_2011}. For planets around both K and G dwarfs, we adopt $F_\mathrm{EUV, 0} = 10^{-3.5} F_\mathrm{p} \left(M_\mathrm{star}/M_\odot \right)$ based on \cite{Jackson_2012} and commonly used in photoevaporation studies \citep[e.g.][]{Owen_2017, Lopez_2018, Rogers_2021}.

In the energy-limited escape regime, EUV-driven escape generates a mass flux [kg\ m$^{-2}$ s$^{-1}$], which we compute as a function of $F_\mathrm{EUV}$, an efficiency factor $\epsilon$, and the planetary gravitational potential $V_\mathrm{pot} = GM_\mathrm{p}/R_\mathrm{p}^2$, where $M_\mathrm{p}$ and $R_\mathrm{p}$ are planetary mass and radius respectively:

\begin{equation}
    \phi_\mathrm{EUV} = \frac{\epsilon F_\mathrm{EUV}}{4 V_\mathrm{pot}}.
    \label{eq:EUVflux}
\end{equation}

\noindent $\epsilon$ encapsulates several heat transfer processes, ultimately representing the fraction of incident radiation that drives escape. We chose a value of $\epsilon$ = 0.15, consistent with previous photoevaporation studies and model estimates \citep{Watson_1981, Owen_2013, Shematovich_2014, Schaefer_2016, Kubyshkina_2018}. As discussed later in Section \ref{sec:deuterium_results}, various cooling mechanisms may lead to $\epsilon$ values below 0.15, one of which we model, as shown in the next paragraph. It may be that radiative cooling becomes more important for higher metallicity atmospheres as fractionation takes place, which should allow more planets to retain fractionated atmospheres. Simulations with $\epsilon$ = 0.05 - 0.10 showed a slight shift in the radius valley, and an accompanying shift in the planet radius-flux space, but still a prominent population of fractionated atmospheres. We leave more detailed modeling of escape efficiency for future work.

A reduction in escape efficiency is expected due to radiative cooling, especially for closer-in planets whose winds are thermostatted at $\approx$ 10,000 K at the $\tau$ = 1 surface so that additional EUV flux does not increase the escape rate \citep{Murray-Clay_2009}. In this case, atmospheric gas is sufficiently ionized such that hydrogen recombination and emission of Ly$\alpha$ photons becomes the dominant cooling mechanism. Planets limited by this cooling mechanism rather than by $F_\mathrm{EUV}$ are in the ``radiative/recombination-limited'' escape regime \citep{Lopez_2018, Lopez_2017, Wordsworth_2018}. Mass flux [kg\ m$^{-2}$ s$^{-1}$] in the radiative/recombination-limited escape regime is computed as:

\begin{equation}
    \phi_\mathrm{RR} = 2 c_\mathrm{s} \sqrt{\frac{F_\mathrm{EUV} \mu_\mathrm{H}^3 g}{h \nu_0 \alpha_\mathrm{rec}^2 k_\mathrm{B} T_\mathrm{eq}}}\  \text{exp} \left[\left(\frac{R_\mathrm{p}}{R_\mathrm{s}} - 1\right) \frac{G M_\mathrm{p}}{R_\mathrm{p} c_\mathrm{s}^2} \right],
    \label{eq:phi_RR}
\end{equation}

\noindent where $c_\mathrm{s} = \sqrt{2 k_\mathrm{B} 10^4/\mu_\mathrm{H}}$ is the sound speed at the sonic point [m/s], $g = G M_\mathrm{p}/R_\mathrm{p}^2$ is the gravitational field strength at the base of the flow [m/s$^2$], $R_\mathrm{p}$ is taken as the minimum of the Bondi radius, Hill radius, and \cite{Lopez_2014} radius prescription [m], $h$ is the Planck constant, $\nu_0 = 4.835\ \text{x}\ 10^{15}$ is the EUV ionizing radiation frequency [Hz; $\approx$60 nm/ 20 eV], $\alpha_\mathrm{rec} = 2.7\ \text{x}\ 10^{-13} (T_\mathrm{eq}/10^4)^{-0.9}/10^6$ is the case B recombination coefficient for hydrogen [m$^3$/atom/s], $k_\mathrm{B}$ is the Boltzmann constant, $T_\mathrm{eq}$ is the planetary equilibrium temperature assuming zero albedo, $R_\mathrm{s} = G M_\mathrm{p}/(2 c_\mathrm{s}^2)$ is the sonic point radius [m], and $M_\mathrm{p}$ is the planet mass [kg]. For all EUV-driven hydrodynamic escape simulations performed with \texttt{IsoFATE}, the mass flux is set by the minimum of $\phi_\mathrm{EUV}$ (Equation \ref{eq:EUVflux}) and $\phi_\mathrm{RR}$ (Equation \ref{eq:phi_RR}).
 
\subsection{Core-powered mass loss} \label{sec:CPML}

Our core-powered mass loss model mirrors that of EUV-driven hydrodynamic escape except that mass loss is powered by the intrinsic planetary luminosity $L_\mathrm{p}$, which results from left over heat of formation rather than by stellar EUV radiation. We follow the prescription outlined in \cite{Gupta_2020}, which assumes the escaping atmosphere is isothermal at $T_\mathrm{eq}$. In the energy-limited escape regime, the mass flux [kg\ m$^{-2}$ s$^{-1}$] expression for the core-powered mass loss mechanism is:

\begin{equation}
    \phi_\mathrm{L} = \frac{\epsilon L_\mathrm{p}}{4 V_\mathrm{pot}},
    \label{eq:phi_L}
\end{equation}

\noindent where $\epsilon$ is a heat transfer efficiency factor and $V_\mathrm{pot}$ is the gravitational potential, as in Equation \ref{eq:EUVflux}. Note that our formulation is equivalent to Equation 9 in \cite{Gupta_2020} with the exception of the added heating efficiency term $\epsilon$, and $L_\mathrm{p}$ is equivalent to their Equation 6. This formulation corresponds to an absolute upper limit on escape rate, assuming all cooling luminosity goes into driving escape.

In the ``Bondi-limited'' escape regime, on the other hand, the escape rate is determined by the thermal velocity of the escaping particles at the Bondi radius. For this case, we again follow the prescription outlined in \cite{Gupta_2020} (Equation 10):

\begin{equation}
    \phi_\mathrm{B} = c_\mathrm{s} \rho_\mathrm{rcb}\ \text{exp} \left[\frac{-G M_\mathrm{p}}{c_\mathrm{s}^2 R_\mathrm{rcb}} \right],
    \label{eq:phi_B}
\end{equation}

\noindent where $\rho_\mathrm{rcb} = 1.0$ kg/m$^3$ is the density at the radiative-convective boundary and $R_\mathrm{rcb}$ is the height of the radiative-convective boundary, calculated as $R_\mathrm{core} + R_\mathrm{env}$ following \cite{Lopez_2014} (i.e. tropopause height). Equation \ref{eq:phi_B} implicitly assumes that the atmosphere is isothermal in the escaping region. For all core-powered mass loss simulations performed with \texttt{IsoFATE}, the mass flux is set by the minimum of $\phi_\mathrm{L}$ (Equation \ref{eq:phi_L}) and $\phi_\mathrm{B}$ (Equation \ref{eq:phi_B}).

\section{Atmospheric Mass Fractionation} \label{sec:isotopic_fractionation}

For a hydrogen-dominated planetary atmosphere undergoing hydrodynamic escape, the primary escaping species is hydrogen. 
The remaining composition of the escaping wind depends on the upward flux of minor species present below the exobase. If the total escape flux is low, H escapes alone. When the escape flux exceeds a critical value, heavier species are dragged along with the escaping flow. Hence, the loss of heavier species is governed by the competition between upward drag by escaping H and downward diffusion by gravity.

Our model simulates this process by numerically integrating via the forward Euler method several differential equations of the general form

\begin{equation}
    \frac{dN_i}{dt} = -A\Phi_i,
    \label{eq:dNdt_main}
\end{equation}

\noindent where $N_i$ is the total number of moles of species $i$, $\Phi_i$ is the number flux of species $i$ [particles m$^{-2}$ s$^{-1}$] defined in Equations \ref{eq:Phi1}, \ref{eq:Phi2}, and \ref{eq:Phi3}, and $A$ is the planetary surface area [m$^2$].

We consider diffusive mass fractionation of escaping atmospheres composed of hydrogen and helium present in Solar abundances \citep{Lodders_2003}. We consider two cases: helium-hydrogen fractionation (He/H) and deuterium-hydrogen fractionation (D/H). For He/H fractionation, we assume a binary mixture where H is the light species and He is the heavy species. We do not expect deuterium to impact this system since it is never the dominant species in the atmosphere. However, for D/H fractionation, we model a ternary mixture of H, D, and He. 

For a binary mixture, we follow the prescription for diffusive fractionation derived in \cite{Wordsworth_2018} \citep[see also][]{Zahnle_1986, Zahnle_1990, Hu_2015}. The number flux [particles m$^{-2}$ s$^{-1}$] of a light species $\Phi_1$ and a heavy species $\Phi_2$ can be calculated as a function of their molar concentrations in the bulk atmosphere ($x_\mathrm{i} = N_\mathrm{i}/N_\mathrm{tot}$), their particle masses $m_\mathrm{i}$, and the total mass flux (from Equation \ref{eq:EUVflux} or \ref{eq:phi_L}; simplified here as $\phi$):

\begin{equation}
    \Phi_1 = 
    \begin{cases}
    \phi/m_1 & \phi < \phi_c \\
    \left[x_1 \phi + x_1 x_2(m_2 - m_1)b_{1,2}/H_2 \right]/ \Bar{m} & \phi \geq \phi_c
    \end{cases}
\label{eq:Phi1}
\end{equation}

and

\begin{equation}
    \Phi_2 = 
    \begin{cases}
    0 & \phi < \phi_c \\
    \left[x_2 \phi + x_1 x_2(m_1 - m_2)b_{1,2}/H_1 \right]/ \Bar{m} & \phi \geq \phi_c,
    \end{cases}
\label{eq:Phi2}
\end{equation}

\noindent where $\Bar{m} = m_1 x_1 + m_2 x_2$ is the mean particle mass of the escaping flow and $b_{1,2}$ is the binary diffusion coefficient for species 1 and 2 and $H_i$ is the effective scale height of species $i$ at the base of the escaping region. We use $b_\mathrm{H,D} = 7.183~\times~10^{19}\ T ^{0.728}$, $b_\mathrm{H,He} = 1.04~\times~10^{20}\ T^{0.732}$, and $b_\mathrm{He,D} = 5.087~\times~10^{19}\ T^{0.728}$ \citep{Mason_1970, Genda_2008}. The quantity $\phi_c$ is the critical mass flux required for the light species to drag the heavier species along with the flow. It is defined as:

\begin{equation}
    \phi_c = \frac{b x_1}{H_1}(m_2 - m_1).
    \label{eq:phicrit}
\end{equation}

This results from setting the two definitions of $\Phi_2$ in Equation \ref{eq:Phi2} equal to each other, setting $\phi = \phi_c$, and using the scale height definition $H_\mathrm{i} = k_\mathrm{B} T/m_\mathrm{i} g$ where $g$ is gravitational field strength, $k_\mathrm{B}$ is the Boltzmann constant, and $T$ is temperature.

Helium enrichment can have an important effect on deuterium fractionation, so we model a ternary mixture of H, D, and He for all D/H fractionation simulations. In this case, we model He/H fractionation as usual with Equations \ref{eq:Phi1} and \ref{eq:Phi2} where species 1 is H and species 2 is He. We derive our D escape flux from an analytical expression for an arbitrary number of species escaping in a subsonic wind from \cite{Zahnle_1990}:

\begin{equation} \label{eq:Z90}
\begin{split}
& \frac{d\ln f_\mathrm{j}}{dr}  = -\frac{GM(m_\mathrm{j}-m_1)}{kTr^2} - \frac{r_0^2}{r^2}\sum_i[\Phi_\mathrm{i} - f_\mathrm{i}\Phi_1 ] \frac{1}{b_\mathrm{i,1}} \\
& + \frac{r_0^2}{r^2}\sum_i[\Phi_\mathrm{i} - \Phi_\mathrm{j}(f_\mathrm{i} /f_\mathrm{j}) ] \frac{1}{b_\mathrm{i,j}},
\end{split}
\end{equation}

\noindent where $f_\mathrm{j} = N_\mathrm{j}/N_1$ is the mixing ratio of species $j$, $N_1$ is moles of the primary escaping species (i.e., H), $r$ is radius, $r_0$ is the planet radius, $M$ is the planet mass, $m$ is atomic mass, $\Phi$ is escape flux as previously defined, and $b_{i,j}$ is the binary diffusion coefficient for species $i$ and $j$.

After setting the total number of species to 3, expanding out sums and solving for $\Phi_3$, we arrive at an expression for the D (species 3) escape flux in a H/He background gas:

\begin{equation}
\Phi_3 = f_3 \frac{\Phi_1 + \alpha_3 \Phi_2 + \alpha_2 \Phi_\mathrm{d,2} x_2 - \Phi_\mathrm{d,3}}{1 + \alpha_3 f_2},
\label{eq:Phi3}
\end{equation}

\noindent where $f_i = x_i/x_1$ (mole fraction), $\alpha_2 \equiv b_\mathrm{H,D}/b_\mathrm{H,He}$ and $\alpha_3 \equiv b_\mathrm{H,D}/b_\mathrm{He,D}$ and $\Phi_{\mathrm{d},i} \equiv b_{1,i}/(H_i^{-1} - H_1^{-1})$. An equivalent expression with a more detailed derivation is reported by \cite{Gu_2023}.

\begin{figure}
    \centering
    \includegraphics[width=\hsize]{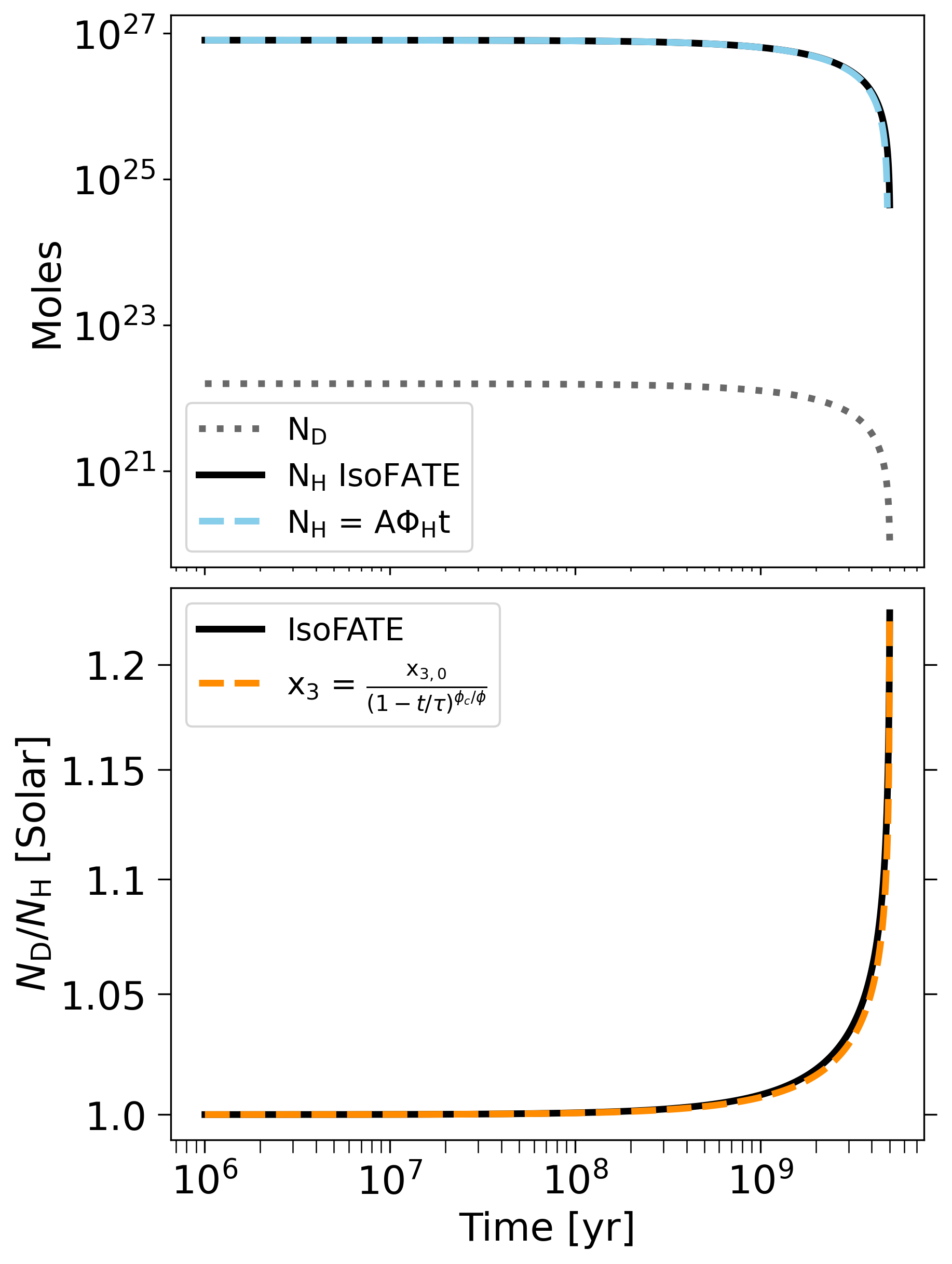}
    \caption{Comparison of our numerical results with an analytic solution. The top panel shows the time evolution of total D and H inventory in moles. The analytic solution for $N_\mathrm{H}$ is represented by the blue dashed line: $A$ is the planetary surface area, $\Phi_\mathrm{H}$ is the H escape flux and $t$ is time. The bottom panel shows the time evolution of the mole fraction of D/H. The analytic solution for $N_\mathrm{D}/N_\mathrm{H}$ is represented by the dashed orange line: x$_3$ is D mole fraction and $\tau = N_\mathrm{H,0}/(A \Phi_\mathrm{H})$ is the atmospheric loss timescale. The derivation is shown in Appendix \ref{sec:analytic}, Equation \ref{eq:analytic1}.}
    \label{fig:analytic_comparison}
\end{figure}

Figure \ref{fig:analytic_comparison} shows a comparison between our numerical model and an analytic solution (derived in Appendix \ref{sec:analytic}) for a representative simulated planet with $M_\mathrm{p}$ = 12 M$_\oplus$, orbital period $P = 10$ days, and initial atmospheric mass fraction $f_\mathrm{atm,0} = 1.5$\%. Unlike our numerical results presented in the following sections, planet radius and escape flux were held constant in this case for simplicity. These results demonstrate that planetary atmospheres experience the most mass fractionation - be it D/H or He/H - toward the end of the escape process, i.e. when nearly the entire atmosphere is lost. Hence, we expect planets that are able to retain a thin gaseous envelope to exhibit the greatest He- and D-enhancement. However, we find that He- and D-enhancement is not solely a function of final atmospheric mass; the escape history is important too. If the escape flux is too high ($\phi >> \phi_c$), lighter species are dragged along with heavier species and fractionation is minimal. Given the canonical interpretation of the radius valley as separating enveloped terrestrial planets from bare rocky cores, we therefore expect planets along the upper radius valley at sufficient orbital separations to exhibit the most fractionation. We validate this prediction in the following section.

\section{Model Simulations} \label{sec:results}

We are interested in understanding the thermally-driven atmospheric escape histories for sub-Neptunes around G, K, and M stars and how their final atmospheric compositions depend on their initial conditions. To explore these questions, we ran a suite of atmospheric evolution models via Monte Carlo simulations over a broad parameter space. For each trial, 500,000 samples were randomly drawn from log uniform grids of initial $M_\mathrm{p}$ between 1 - 20 M$_\oplus$, initial $f_\mathrm{atm}$ between 0.1 - 50\%, and $P$ between 1 - 300 days. Atmospheric mass loss was initiated at 1 Myr. Orbital migration effects were ignored so $P$ was held constant. Model simulations were halted at 5 Gyr. We assume Solar composition for initial D/H and He/H values according to \cite{Lodders_2003}, though in reality these values are expected to deviate for different systems. An average trial of 500,000 simulated planets runs for approximately 36 hours, affording rapid exploration of a broad parameter space to yield population-level predictions. We focus on the results for planets around low-mass stars, given their greater observability. We also choose to simulate escape with EUV-driven photoevaporation alone, core-powered mass loss alone, and the combination of the two. There is a great body of empirical evidence (Section \ref{sec:intro}) to suggest that these mechanisms may be sculpting the radius distribution around G, K, and M stars. More recent work suggests that each may operate independently or in tandem for sub-Neptunes \citep{Owen_2023}.

In order to test our prediction of fractionation along the upper radius valley and to compare our model to previous studies, we determined the location of the radius valley in our simulations by fitting a linear model to log$R_\mathrm{p}$ vs. log$P$ following the approach taken by \cite{Van_Eylen_2018}. This approach uses support vector machines to determine the hyperplane of maximum separation between planets above and below the radius valley, which is a line in this case. For this purpose, we classify planets below the valley as ``super-Earths," defined as having lost their entire atmospheres, while planets above the valley are classified as ``sub-Neptunes" and have retained a H/He envelope. The line of separation maximizes the distance to points of these two pre-defined classes of simulated data. To obtain the model fits, we employed the Python support vector classification routine in the {\tt\string scikit} machine learning package {\tt\string scikit-learn}. This method requires choice of a penalty parameter C, which sets a tolerance for misclassification, with high values allowing lower misclassification. We tested values between C = 0.1 - 100 and determined that C = 100 provided the best visual fit to the data and best agreed with previously reported values. Larger values of C resulted in equally good fits.

\begin{figure*}
    \centering
    \includegraphics[width=0.75\textwidth]{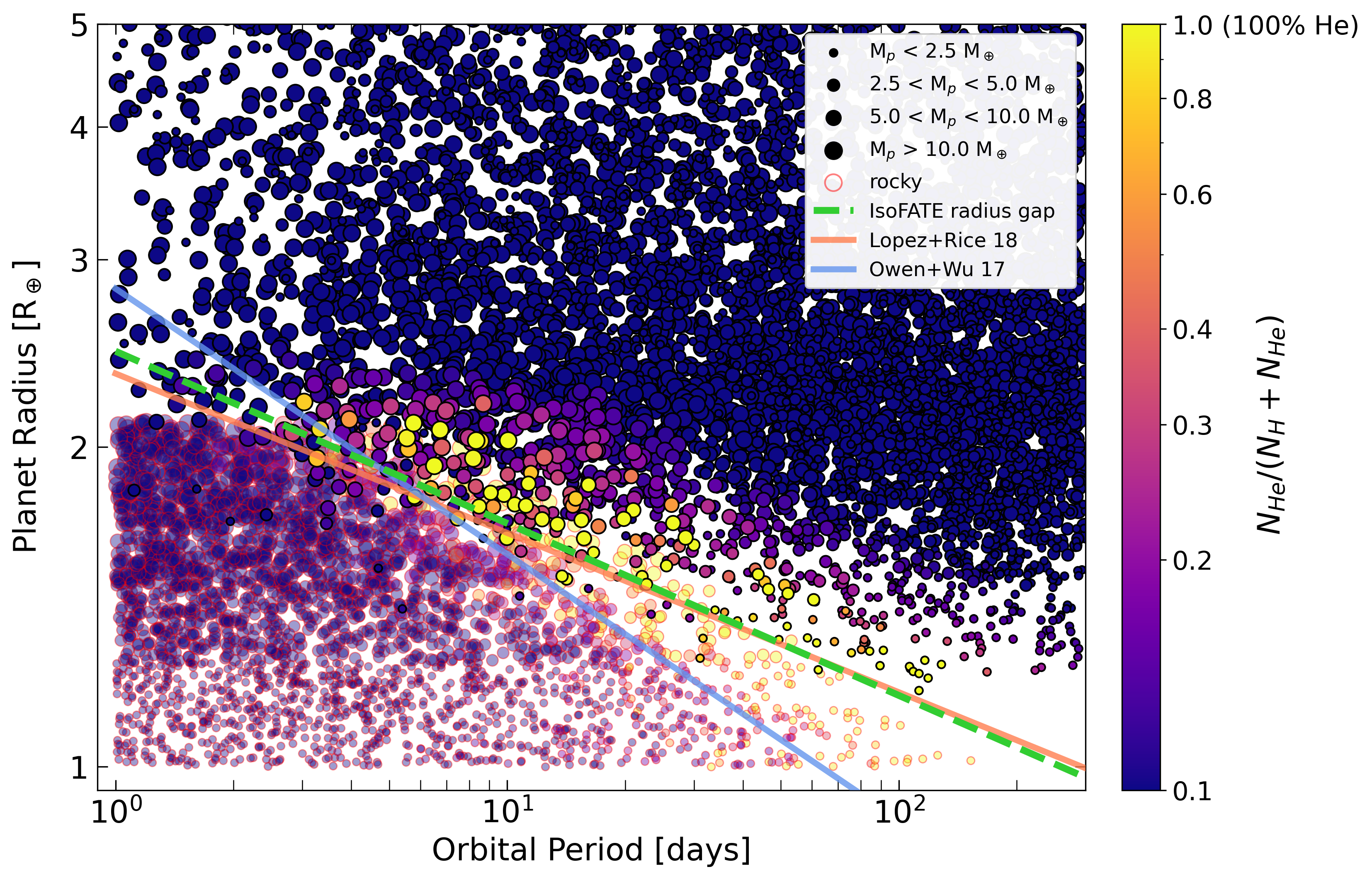}
    \caption{EUV-driven photoevaporation simulations for a random subset of n = $1 \times 10^4$ planets around an early M dwarf. Helium molar concentration is displayed as a function of orbital period and final planetary radius after 5 Gyr. Planets that lost their entire atmospheres are circled in red and labeled ``rocky''. Marker size depicts planet mass. Planets along the upper edge of the radius valley undergo the most fractionation.}
    \label{fig:He_mc}
\end{figure*}

\subsection{Helium Enhancement} \label{sec:helium_results}

\begin{figure}
    \centering
    \includegraphics[width=\hsize]{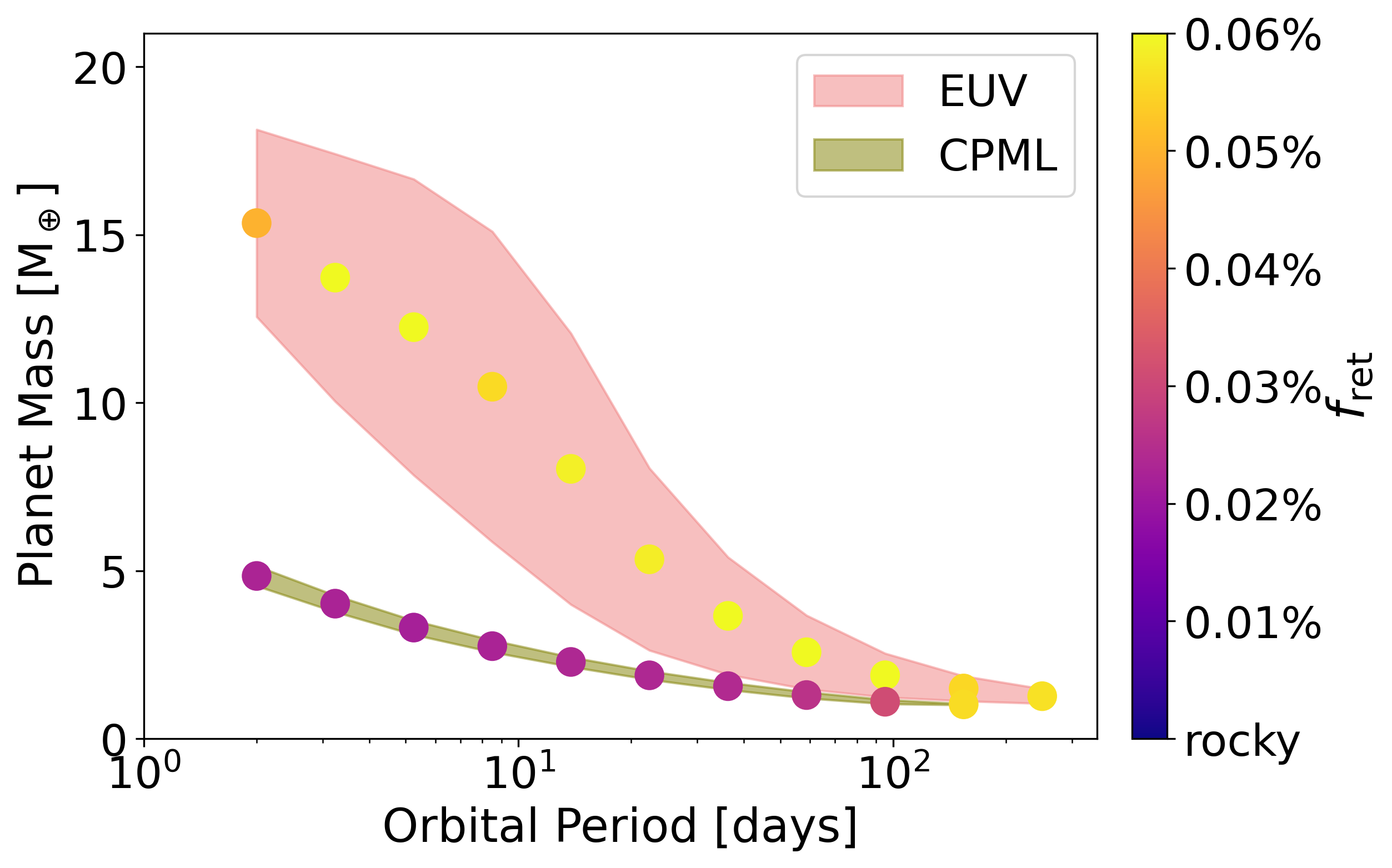}
    \caption{Planet mass and final atmospheric mass fraction, $f_\mathrm{ret}$, as a function of orbital period for the subset of simulated planets with helium-enriched atmospheres ($x_\mathrm{He} \geq 0.145$) for both loss mechanisms. The shaded areas represent the standard deviation. Mass is negatively correlated with orbital period for both mechanisms. Highly fractionated planets retain larger atmospheres and are generally more massive in the EUV-driven photoevaporation simulations.}
    \label{fig:mass_period}
\end{figure}

Following \cite{Hu_2015}, \cite{Malsky_2020}, and \cite{Malsky_2023}, we first explored helium enhancement of sub-Neptune atmospheres through hydrodynamic escape. Figure \ref{fig:He_mc} shows a set of simulations for planets around an early M dwarf experiencing EUV-driven photoevaporation. As expected, significant helium enhancement is seen for planets along the upper edge of the radius valley. For planets with high fractionation, there is a negative correlation between orbital period and planetary mass, which is further highlighted in Figure \ref{fig:mass_period}. Figure \ref{fig:He_mc} also shows reasonable agreement of the radius valley location with \cite{Lopez_2018} and \cite{Owen_2017}. Note that only the radius valley slope is reported in these works and the vertical offset is approximated by eye here.

In order to maximize fractionation for a given planetary atmosphere, the condition $\phi \approx \phi_c$ must be met so that a sufficient quantity of the lighter species is driven away and a sufficient inventory of the heavier species is retained. For $\phi >> \phi_c$, the heavier species fails to diffuse downward through the escaping flow rapidly enough and is dragged along with the lighter escaping species. Hotter, more irradiated planets closer to their host stars must possess stronger gravitational fields and hence be more massive in order to undergo fractionation and avoid excessive escape fluxes. This critical planetary mass decreases with increasing orbital period, hence the negative correlation seen in Figures \ref{fig:He_mc} and \ref{fig:mass_period}. The correlation is stronger for EUV-driven photoevaporation since it is a stronger driver of escape than core-powered mass loss (hereafter ``CPML''), leading to more massive fractionated planets as seen in Figure \ref{fig:mass_period}.

\begin{figure*}
    \centering
    \includegraphics[width=\textwidth]{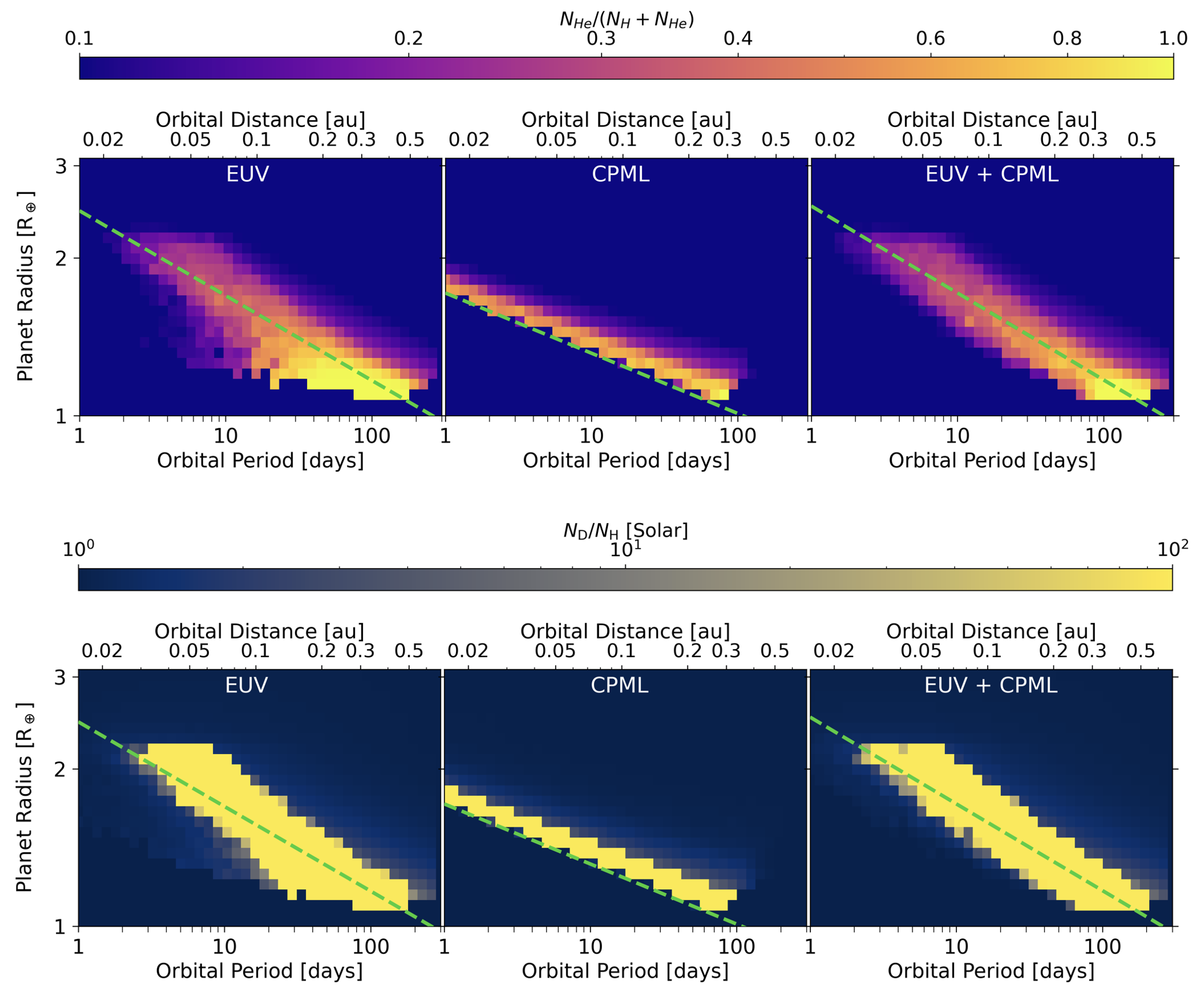}
    \caption{Helium molar concentration (top panel) and deuterium mole fraction (bottom panel) as a function of orbital period and planet radius for EUV-driven photoevaporation (left) core-powered mass loss (middle) and the combined mechanisms (right). These plots are derived from simulations with n = 5 x 10$^5$ planets around M dwarfs, like that shown in Figure \ref{fig:He_mc}. The mean He (top) or D (bottom) abundance of all sub-Neptunes in each grid cell that have retained atmospheres after 5 Gyr is shown. The dashed green line represents the calculated radius valley as described in Section \ref{sec:results}. Significant helium and deuterium enhancement is observed along the radius valley in all cases.}
    \label{fig:PR_contours}
\end{figure*}

The top panel of Figure \ref{fig:PR_contours} shows helium molar concentrations for EUV-driven photoevaporation (left), CPML (middle), and the combined mechanisms (right) (Compare to \cite{Malsky_2020} Figure 5 and \cite{Malsky_2023} Figure 1). Qualitatively, our results resemble those of Malsky et al., with significant helium enrichment observed along the upper edge of the radius valley. However our results differ in that the area near the radius valley is more populated with fractionated planets, whereas the results of Malksy et al. show significant gaps in the distribution near the radius valley, presumably due to the reported failure of the MESA code's equations of state for planets with tenuous atmospheres. Another difference is that the most fractionated planets appear at greater orbital distances in our results while they appear at intermediate orbital distances in the results of Malsky et al. Finally, while Malsky et al. report planets with atmospheres $\gtrsim$ 80\% helium by mass, our simulations yield planets with greater overall helium enhancement, reaching $\approx$ 100\%.

Figure \ref{fig:PR_contours} was generated by binning data from simulations represented in Figure \ref{fig:He_mc}. The mean helium molar concentration is computed for each grid cell only for planets that have retained an atmosphere at the end of the simulation. We observe closer-in helium-enhanced planets ($P \lesssim 10$ days) have greater planetary radii in the EUV-driven photoevaporation simulations compared to those in the CPML simulations. This follows from Figure \ref{fig:mass_period} which shows that these planets are more massive and possess greater final atmospheric mass fractions in the EUV-driven case compared to the CPML case. This is also reflected in the steeper radius valley reported in previous photoevaporation simulations compared to that in CPML simulations \citep{Lopez_2018, Gupta_2020}. \cite{Malsky_2023} report similar findings for EUV-driven photoevaporative mass loss, with planets in their simulations achieving He mass fractions $\geq 0.40$ after 2.5 Gyr and $\geq 0.80$ after 10 Gyr. The different parameter spaces occupied by helium-enhanced planets observed in this study offers a novel approach to investigate whether EUV-driven photoevaporation or CPML dominates sub-Neptune atmosphere evolution around low-mass stars. We discuss observational prospects to this end in Section \ref{sec:Observation}.

\subsection{Deuterium Enhancement} \label{sec:deuterium_results}

We next explore the atmospheric fractionation of deuterium and hydrogen (i.e., protium) as a result of hydrodynamic escape. Previous work by \cite{Gu_2023} report modest D/H fractionation for sub-Neptunes around Solar-type stars (max $N_\mathrm{D}/N_\mathrm{H} \approx 1.7$x Solar). Like the He fractionation studies of \cite{Malsky_2020} and \cite{Malsky_2023}, \cite{Gu_2023} utilized the MESA code and encountered anomalies in the data for small planet radii. This shortcoming prevented the smooth transition to a bare rocky core as a planet lost its atmosphere. As we expect from analytic solutions (e.g. Figure \ref{fig:analytic_comparison}) and as we see from our simulations, planetary atmospheres are significantly fractionated in the final stages of losing their atmospheres. The loss process is imprinted on the planet only if the planet manages to retain a gaseous envelope. Hence, previous work that employed the MESA code to compute equations of state was unable to probe the parameter space in which planets become deuterium-enhanced.

Here we build on these previous studies and explore a wider parameter space, to provide a more complete picture of deuterium fractionation. The bottom panel of Figure \ref{fig:PR_contours} shows deuterium mole fraction as a function of orbital period and planet radius (Compare to Figure 2 in \cite{Gu_2023}). Note that the D/H mole fraction is capped at 100x Solar in Figure \ref{fig:PR_contours} and many planets retain atmospheres with essentially no hydrogen remaining, only deuterium. Though qualitatively similar to past work, our results demonstrate that deuterium enhancement was previously dramatically underestimated. We observe significant fractionation along the upper radius valley for all three mechanism combinations in our simulated data. Not surprisingly, we observe deuterium-enriched planets in the same parameter space as that of helium-enriched planets. In fact, we find that 100\% of deuterium-enriched planets are also helium enriched.

\begin{figure}
   \centering
  \includegraphics[width=\hsize]{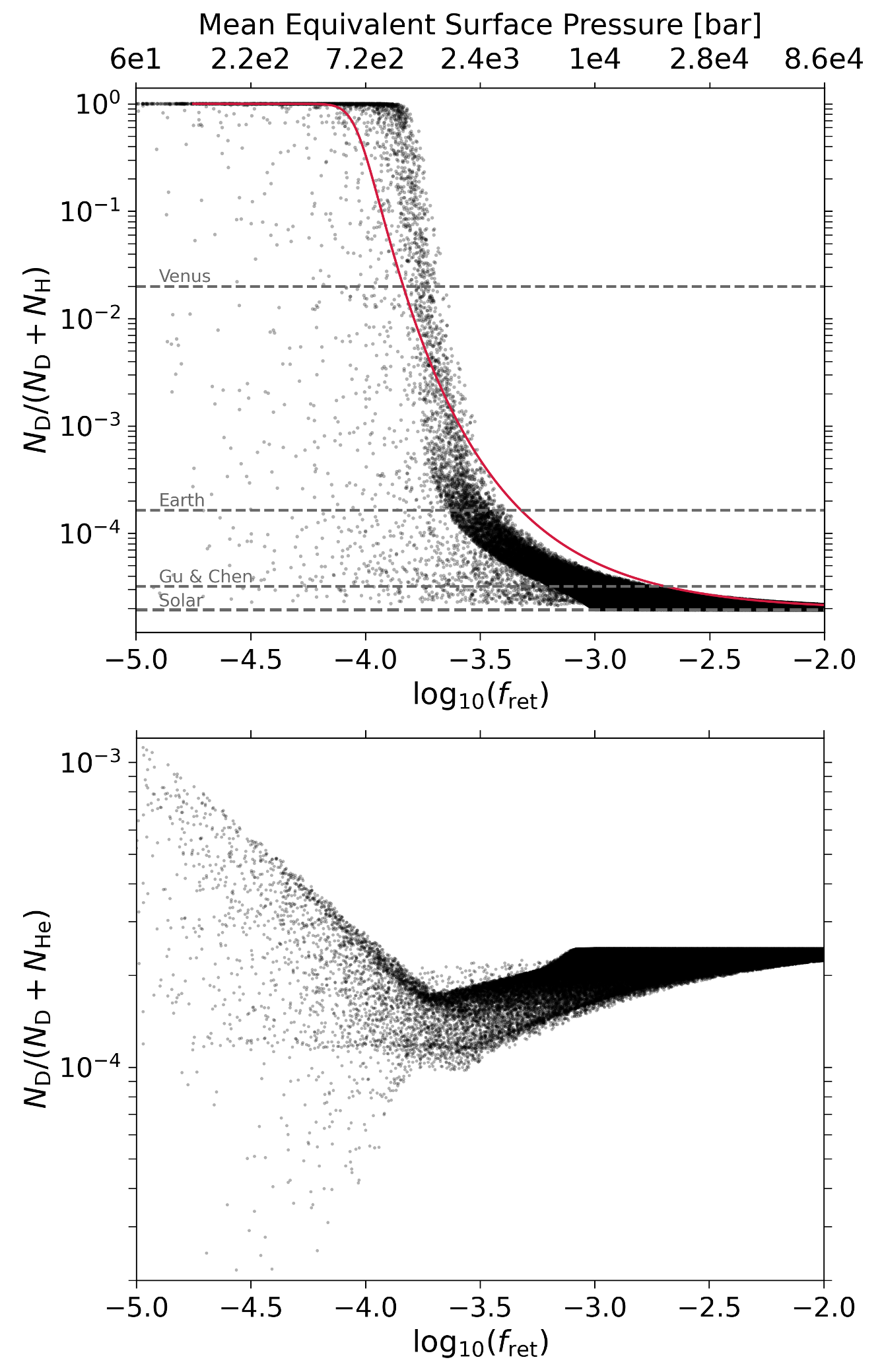}
   \caption{Molar concentrations of deuterium relative to H (top) and He (bottom) as a function of retained atmospheric mass fraction, $f_\mathrm{ret}$, for EUV-driven photoevaporation. The upper axis corresponds to the mean surface pressure for each $f_\mathrm{ret}$ bin. Simulated data correspond to the final state for $5 \times 10^5$ simulations of planets around an early M dwarf evolved for 5 Gyr. The red line shows an analytic solution for molar concentration of D as a function of $f_\mathrm{ret}$ and $T_\mathrm{eq}$ (derived in Appendix \ref{sec:analytic2}). The mean $T_\mathrm{eq}$ = 500 K was used in this case. Venus and Earth dashed lines show deuterium concentrations corresponding to Venusian atmospheric D/H and the Vienna Standard Mean Ocean Water (VSMOW) values respectively. The ``Gu \& Chen'' dashed line shows the maximum D/H value reported by \cite{Gu_2023} in their photoevaporation simulations. The bottom panel shows that most planets contain atmospheric deuterium in the hundreds of ppm.}
   \label{fig:mean_DH}
\end{figure}

Our results suggest that atmospheric D/H may serve as a useful diagnostic of surface pressure for sub-Neptunes with thin hydrogen/helium envelopes. Figure \ref{fig:mean_DH} shows deuterium concentration as a function of final atmospheric mass fraction, $f_\mathrm{ret}$, for planets around an M dwarf experiencing EUV photoevaporation. We observe a sharp transition between $f_\mathrm{ret} = 10^{-4}$ - $10^{-3.5}$ where planets essentially lose all of their hydrogen, and only trace deuterium in a thin, helium-dominated atmosphere remains (surface pressure $\lesssim$ 1,000 bar). Differences in results for other stellar types were negligible, though the curve is shifted to lower $f_\mathrm{ret}$ values by about 0.5 dex for the CPML mechanism. As we have demonstrated, the most tenuous atmospheres tend to be the most fractionated. Hence, atmospheric D/H can serve as a proxy for the upper limit of surface pressure. \cite{Misener_2021} propose that the core-powered mass loss mechanism may shut down when the cooling timescale equals the loss timescale, leaving behind a tenuous atmosphere. Other factors such as molecular line cooling may decrease escape efficiency, similarly resulting in retention of tenuous atmospheres \citep{Nakayama_2022}. \cite{Misener_2021} report analytical estimates of $f_\mathrm{ret} \in [10^{-4}, 10^{-8}]$ which overlap with values corresponding to elevated D/H in our simulations (see their Figure 5). The $f_\mathrm{ret}$ values of $10^{-3},\ 10^{-4},$ and $10^{-5}$ correspond to mean surface pressures, $P_\mathrm{surface}$, of approximately 10,000, 700, and 60 bar respectively in our simulations, as seen in Figure \ref{fig:mean_DH}. These values are calculated as the mean $P_\mathrm{surface}$ for all simulated planets within $f_\mathrm{ret} \pm 0.5$ dex, where $P_\mathrm{surface} = G M_\mathrm{p}^2 f_\mathrm{ret}/ 4 \pi R_\mathrm{core}^4$. Since D-enriched atmospheres tend to be He-dominated, a similar relationship is observed for $f_\mathrm{ret}$ and $x_\mathrm{He}$ suggesting He/H can be used as a proxy for $P_\mathrm{surface}$ too.

While constraining atmospheric D/H presents an observational challenge, depletion of hydrogen in deuterium-enhanced, helium-dominated atmospheres may confer an advantage. \cite{Molliere_2019} show that HDO may be detectable in sub-Neptune atmospheres with D/H as low as the VSMOW value, so long as methane (CH$_4$) is depleted to leave the 3.7 $\mu$m window open. Fortuitously, deuterium-enhanced planets are typically helium-dominated and are expected to be depleted in CH$_4$. In a solar composition atmosphere $\lesssim$ 700 K, the dominant carbon-bearing species is CH$_4$ \citep{Fortney_2020}. For an evolved atmosphere, CH$_4$ is depleted in favor of CO and CO$_2$ when $x_\mathrm{H} < x_\mathrm{C} + x_\mathrm{O}$ \citep{Hu_2015}, a condition commonly achieved in our simulations using He as a proxy for heavier species. Therefore, a residual helium atmosphere will have most of its carbon inventory in CO and CO$_2$ rather than CH$_4$, making HDO more observable and D/H measurement more feasible.


Our calculations assume that the planet forms with a rocky core and a solar-composition atmosphere. Water world exoplanets, which may have significant fractions of their mass as H$_2$O, would likely exhibit much less D/H fractionation due to rapid H and D exchange between H$_2$O reservoirs and atmospheric H$_2$, unless they also lost all of their H$_2$O inventory \citep{Genda_2008}. Hence D/H observations should also provide a way to distinguish water worlds from fractionated sub-Neptunes.

\section{Observational Prospects} \label{sec:Observation}

\begin{deluxetable*}{lccccccp{3cm}p{3cm}}
\tabletypesize{\small}
\tablecaption{Fractionated planet candidates predicted from $M_\mathrm{p}$-$R_\mathrm{p}$-$F_\mathrm{p}$ interpolation \label{tab:targets}}
\tablehead{\colhead{Planet} & \colhead{Stellar type} & \colhead{M$_\mathrm{p}$ [M$_\oplus$]} 
& \colhead{R$_\mathrm{p}$ [R$_\oplus$]} & \colhead{P [days]} & \colhead{J-band mag} & \colhead{TSM} & \colhead{\textbf{He/H} Mechanisms ($P_\mathrm{frac}$)} & \colhead{\textbf{D/H} Mechanisms ($P_\mathrm{frac}$)}}
\startdata
\\ 
LHS 1140 c & M & 1.91 & 1.272 & 3.78 & 9.61 & 147 & EUV (1.0) \newline EUV+CPML (1.0) & EUV (1.0) \newline EUV+CPML (1.0) \\
L 98-59 d & M & 2.31 & 1.58 & 7.45 & 7.93 & 236 & EUV (0.16) \newline EUV+CPML (0.24) & None \\
TOI-1452 b & M & 4.82 & 1.67 & 11.06 & 10.60 & 39 & EUV (0.47) \newline EUV+CPML (0.46) & None \\
Kepler-138 d & M & 2.10 & 1.51 & 23.09 & 10.29 & 23 & EUV (0.26) \newline CPML (0.12) \newline EUV+CPML (0.24) & None \\
Kepler-138 c & M & 2.30 & 1.51 & 13.78 & 10.29 & 25 & EUV (0.33) \newline CPML (0.26) \newline EUV+CPML (0.36) & None \\
K2-3 c & M & 2.68 & 1.58 & 24.65 & 9.42 & 30 & EUV (0.21) \newline CPML (0.14) \newline EUV+CPML (0.22) & None \\
HD 260655 c & M & 3.09 & 1.53 & 5.71 & 6.67 & 196 & EUV (0.54) \newline CPML (0.49) \newline EUV+CPML (0.57) & None \\
TOI-1695 b & M & 6.36 & 1.90 & 3.13 & 9.64 & 42 & EUV (0.11) \newline CPML (0.07) \newline EUV+CPML (0.12) & None \\
\\ 
Kepler-48 b & K & 3.94 & 1.88 & 4.78 & 11.70 & 13 & EUV (0.04) & None \\
Kepler-161 b & K & 12.1 & 2.12 & 4.92 & 12.77 & 4 & EUV (0.07) \newline EUV+CPML (0.07) & None \\
Kepler-48 d & K & 7.93 & 2.04 & 42.90 & 11.70 & 4 & EUV (0.06) \newline CPML (0.03) \newline EUV+CPML (0.05) & None \\
HD 23472 b & K & 8.32 & 2.00 & 17.67 & 7.87 & 36 & EUV (0.09) \newline CPML (0.10) \newline CPML+EUV (0.09) & None \\
Kepler-80 e & K & 4.13 & 1.60 & 4.64 & 12.95 & 6 & CPML (0.42) & EUV (0.35) \newline CPML (0.16) \newline EUV+CPML (0.34) \\
TOI-836 b & K & 4.53 & 1.70 & 3.82 & 7.58 & 82 & CPML (0.32) & None \\
TOI-178 c & K & 4.77 & 1.67 & 3.24 & 9.37 & 34 & CPML (0.27) & EUV (0.34) \newline CPML (0.13) \newline EUV+CPML (0.30) \\
K2-199 b & K & 6.90 & 1.73 & 3.23 & 10.28 & 16 & None & EUV (0.38) \newline EUV+CPML (0.42) \\
\\ 
Kepler-538 b & G & 12.90 & 2.22 & 81.74 & 10.03 & 3 & EUV (0.06) \newline EUV+CPML (0.06) & None \\ 
\enddata
\tablecomments{Planet parameters obtained from the \dataset[NASA Exoplanet Archive]{\doi{10.26133/NEA13}}, queried on August 24, 2023. TSM calculation follows prescription in \cite{Kempton_2018}.
}
\end{deluxetable*}

\begin{figure*}
    \centering
    \includegraphics[width=\textwidth]{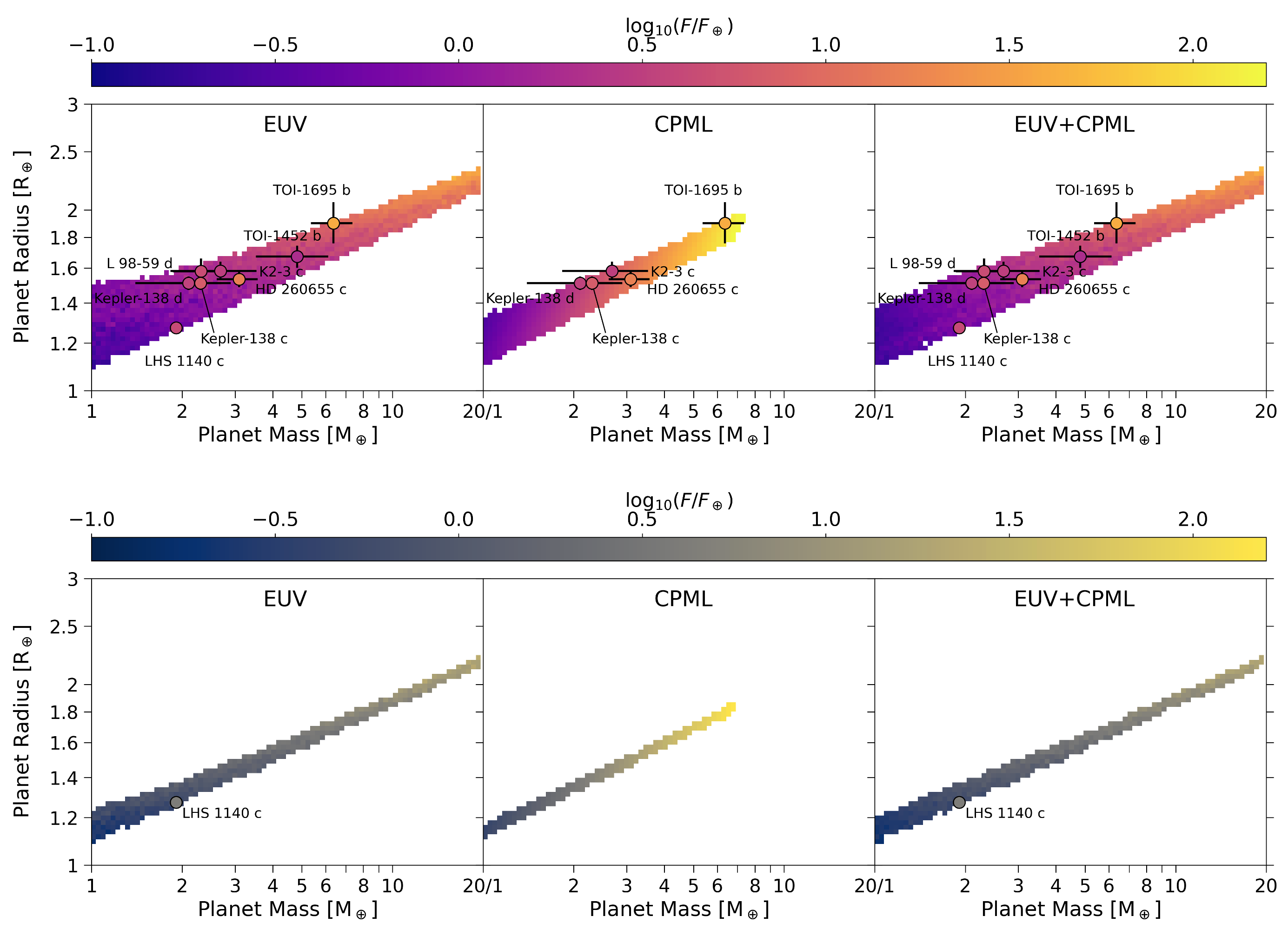}
    \caption{Masses and radii for the subset of simulated planets with helium-enriched atmospheres ($x_\mathrm{He} \geq 0.145$; top panel) and deuterium-enriched planets ($N_\mathrm{D}/N_\mathrm{H} \geq 100$x Solar; bottom panel) overlaid with planets around M dwarfs. Colors represent the mean incident planetary flux for all simulated planets in each grid cell. We include planet candidates whose calculated flux is within ± 1 dex of the interpolated flux to account for uncertainty in instellation history.}
    \label{fig:MR_contours}
\end{figure*}

To identify planet candidates with helium- and deuterium-enriched atmospheres, we queried the NASA Exoplanet Archive on August 24, 2023 for sub-Neptunes in the relevant parameter space. We filtered the sub-Neptunes for those with mass and radius measurement precision of at least 55\% and 8\% respectively and mass-radius profiles inconsistent with purely rocky compositions. We then filtered our simulated data for those planets with helium molar concentration of $x_\mathrm{He}\geq 0.145$ (equivalent to \cite{Malsky_2023} cut off at mass fraction = 0.4) and deuterium mole fraction of $N_\mathrm{D}/N_\mathrm{H} \geq$ 100x Solar, motivated by deuterium isotopologue observability predictions with JWST and high resolution ground-based spectroscopy for terrestrial exoplanets \citep{Lincowski_2019, Molliere_2019}. Using our simulated data, we created a three-dimensional grid of planet mass, planet radius, and planet incident flux. The mean planet flux was calculated for all fractionated planets in each mass-radius grid cell (Figure \ref{fig:MR_contours}). Finally, we used the {\tt\string RegularGridInterpolator} routine in the {\tt\string SciPy} Python library to interpolate the flux at the mass-radius location of our planet candidates. We have demonstrated that atmospheric fractionation is largely a function of planet mass, planet radius, and stellar irradiation history. Of these parameters, irradiation history is the most uncertain for the population of exoplanets with measured masses and radii. Our analysis ignores orbital migration, short-term stellar activity, planet albedo, and other factors affecting irradiation history. Hence we consider a range of flux values in which our planet candidates could fall to be considered an observational prospect. Following from \cite{Malsky_2023}, we compared the interpolated flux to the flux calculated for each planet. If the interpolated flux fell within 0.1x and 10x the calculated flux, then we included the planet in our list. The results are presented in Figure \ref{fig:MR_contours} and Table \ref{tab:targets}. The rightmost columns in Table \ref{tab:targets} show the mechanisms for which He/H and D/H fractionation are predicted along with an estimate of the probability of fractionation for each target, $P_\mathrm{frac}$. This probability is the averaged number of fractionated planets divided by the total number of planets in each grid cell occupied by the target and its error bars in $M_\mathrm{p}$-$R_\mathrm{p}$-$F_\mathrm{p}$ space (Figure \ref{fig:MR_contours}). The transmission spectroscopy metric (TSM) was calculated following the prescription outlined in \cite{Kempton_2018}.

The parameter space populated by simulated, fractionated planets is visibly different between mechanisms. That of the CPML simulations particularly stands out. As seen in Figure \ref{fig:mass_period}, planets undergoing CPML alone have lower masses than their counterparts undergoing EUV photoevaporation. These planets also have lower $f_\mathrm{ret}$ values ($\approx 0.02\%$). A narrow distribution of $f_\mathrm{ret}$ values will produce a narrow distribution of planetary radii. Hence we see a narrower band of fractionated planets in $M_\mathrm{p}$-$R_\mathrm{p}$ space for the CPML scenario. We see wider bands for planets undergoing EUV photoevaporation because the additional parameter of variable EUV flux over time results in a wider distribution of allowed planet masses and initial atmospheric mass fractions that result in residual fractionated atmospheres. We also note that in the EUV scenarios, the maximum planetary flux is lower than that in the CPML scenario for a given planetary mass because atmospheres of close-in planets cannot survive the intense irradiation which results in complete stripping of the atmosphere.

Since the $M_\mathrm{p}$-$R_\mathrm{p}$-$F_\mathrm{p}$ parameter space is different for each mechanism combination, the planets predicted to have fractionated atmospheres differ by mechanism as well (See Table \ref{tab:targets}). Planets around K dwarfs are particularly interesting since they are especially amenable to helium detection via the 1083 nm line \citep{Oklopcic_2019}. Three targets predicted to have helium atmospheres stand out among this group: Kepler-80 e, TOI-836 b, and TOI-178 c. These three targets are predicted to have helium enhancement in the CPML scenario, but not when EUV photoevaporation is included. These planets offer a unique means to test which thermally-driven escape mechanism dominates the sculpting of planets around K dwarfs. Of these targets, TOI-836 b and TOI-178 c have been allocated time on JWST for atmospheric characterization \citep{Batalha_2021, Hooton_2021}. Kepler-80 e has a TSM of 6, and is likely unobservable with current facilities.

The remaining targets around K dwarfs and those predicted to be fractionated around M dwarfs offer a means to test if thermally-driven escape mechanisms work to sculpt the radius valley versus alternative hypotheses such as gas-depleted formation \citep{Lee_2014, Lopez_2018, Lee_2021}. Recent work has demonstrated that the radius valley for planets around low mass stars may result from a unique channel of planet formation entirely different than the thermally-driven escape processes evidenced to be sculpting the radius valley of planets around Sun-like stars \citep{Cloutier_2020, Cherubim_2023}. Here we propose a new strategy for investigating this open question through constraining He/H and D/H ratios of planets around low mass stars.

\section{Conclusion} \label{sec:Conclusion}

\begin{enumerate}
    \item Helium-dominated, deuterium-rich atmospheres are predicted for sub-Neptunes along the upper edge of the radius valley for planets around G, K, and M stars. This novel class of planets spans a large range of $T_\mathrm{eq} \in$ [150 K, 890 K], with a mean of 370 K, largely independent of stellar type, placing some planets in their habitable zones. Closer-in planets require a greater mass to retain fractionated atmospheres in the face of enhanced escape. This is especially true for EUV photoevaporation.


    \item Atmospheric fractionation is mechanism-dependent. We show that EUV-driven photoevaporation fractionates planets in a wider planet/atmospheric mass distribution relative to CPML, endowing them with a wider distribution of final atmospheric mass fraction ($f_\mathrm{ret, EUV} \approx 0.06 \%$, $f_\mathrm{ret, CPML} \approx 0.03 \%$). More massive planets tend to become fractionated via EUV-driven photoevaporation compared to CPML.
    
    \item Our numerical model {\tt\string IsoFATE} overcomes challenges presented by the MESA model used in previous studies by \cite{Malsky_2020}, \cite{Malsky_2023}, and \cite{Gu_2023} that preclude resolution of atmospheric composition for planets with the most tenuous atmospheres, which are the ones that undergo the most fractionation. Hence we expand the parameter space for helium and deuterium enhancement predictions and demonstrate that fractionation was previously underestimated.

    \item Constraining atmospheric He/H and D/H abundance for sub-Neptunes offers a unique means of investigating competing mechanisms that may explain the origin of the radius valley for low mass stars. Observational studies are needed to test existing model predictions. We leave more detailed modeling of processes that may confound the fractionation/escape relationship, such as cometary delivery and interior-atmosphere exchange, for future studies.

    \item Helium-dominated atmospheres are expected to be depleted in CH$_4$ and instead possess CO$_2$ \citep{Hu_2015}. This leaves the 3.7 $\mu$m spectral window open to detect HDO, a novel observing strategy potentially feasible for even modest D/H values with high-resolution ground-based spectroscopy \citep{Molliere_2019}.
\end{enumerate}

\software{{\tt scikit-learn} \citep{scikit-learn},
            {\tt SciPy} \citep{scipy}
          }

\section*{Acknowledgments}

We thank Omar Jatoi for his support in parallelizing the {\tt\string IsoFATE} code, a requisite for efficiently performing Monte Carlo simulations that made this study possible.

This research has made use of the NASA Exoplanet Archive, which is operated by the California Institute of Technology, under contract with the National Aeronautics and Space Administration under the Exoplanet Exploration Program. The data can be accessed via \dataset[NASA Exoplanet Archive]{\doi{10.26133/NEA13}}.

RW acknowledges funding from NSF-CAREER award AST-1847120 and Virtual Planetary
Laboratory (VPL) award UWSC10439.


\bibliographystyle{aasjournal}
\bibliography{refs.bib}

\appendix

\section{Analytic solution of D/H in a binary mixture} \label{sec:analytic}
We present an analytic solution for species abundance in a binary mixture to use for model validation. The solution is compared to our numerical model in Figure \ref{fig:analytic_comparison}. 

Consider a binary mixture of atmospheric species 1 and 3 where $m_1 < m_3$. For species $i$, the total number of moles is $N_i$ and the time derivative of $N_i$ is Equation \ref{eq:dNdt_main}:

\begin{equation}
    \frac{dN_i}{dt} = -A\Phi_i,
    \label{eq:dNdt}
\end{equation}

\noindent where $A$ is surface area. The molar concentration of species 3 is $x_3 = N_3/(N_1 + N_3)$. For H as species 1 and D as species 3, $x_3 \approx N_3/N_1 << 1$, giving

\begin{equation}
    \frac{dx_3}{dt} \approx \frac{1}{N_1} \left( \frac{dN_3}{dt} - \frac{N_3}{N_1} \frac{dN_1}{dt} \right) = \frac{1}{N_1} \left(\frac{dN_3}{dt} - x_3 \frac{dN_1}{dt} \right).
    \label{eq:dx2dt}
\end{equation}

\noindent By definition $x_1 + x_3 = 1$, so $x_1 \approx 1$. Since $x_3 << 1$, $\Bar{m} \approx m_1$, and we get:

\begin{equation}
    \frac{dx_3}{dt} = \frac{1}{N_1} \left( -A \Phi_3 + x_3 A \Phi_1 \right) = \frac{A}{m_1 N_1} \left(-m_1 \Phi_3 + x_3 m_1 \Phi_1 \right).
    \label{eq:dx2dt2}
\end{equation}

\noindent The species escape fluxes, $\Phi_1$ and $\Phi_3$, are represented by the second lines in Equations \ref{eq:Phi1} and \ref{eq:Phi2} when they exceed the critical flux as defined by Equation \ref{eq:phicrit}. Substituting $b_{1,3}/H_3 = (m_3/m_1) b_{1,3}/H_1$ into Equation \ref{eq:Phi1}, we get

\begin{equation}
    \Phi_3 = \left[x_1 \phi + x_1 x_3(m_3 - m_1)\frac{m_3}{m_1}\frac{b_{1,3}}{H_1} \right]/ \Bar{m}
    \label{eq:Phi1_app}
\end{equation}

\noindent and

\begin{equation}
    \Phi_3 = \left[x_3 \phi + x_1 x_3(m_1 - m_3)\frac{b_{1,3}}{H_1} \right]/ \Bar{m}
    \label{eq:Phi2_app}
\end{equation}

\noindent for species 1 and 3 respectively. Substituting Equations \ref{eq:Phi1_app} and \ref{eq:Phi2_app} into Equation \ref{eq:dx2dt2}, we get

\begin{equation}
    \frac{dx_3}{dt} = \frac{A x_3}{m_1 N_1} (m_3 - m_1) \frac{b_{1,3}}{H_1}.
\end{equation}

\noindent This simplifies to

\begin{equation}
    \frac{d \mathrm{log} x_3}{dt} \approx \frac{A}{N_1} \left( \frac{m_3}{m_1} - 1 \right) \frac{b_{1,3}}{H_1}.
\end{equation}

\noindent In the special case where $\Phi_1$ is constant and $N_1(t) = N_1(0) - A\Phi_1 t$ until $t = \tau$ where $\tau = N_1(0)/A \Phi_1$ is the loss timescale, we get

\begin{equation}
    \frac{d \mathrm{log} x_3}{d(t/\tau)} \approx \frac{\gamma}{1 - t/\tau},
\end{equation}

\noindent where $\gamma \equiv (m_3/m_1 - 1)b_{1,3}/({H_1}\Phi_1)$. If we hold the gravitational field strength and temperature constant, then $\Phi_\mathrm{d,1}$ is also constant and we can integrate to find

\begin{equation}
    x_3 = x_3(0) \frac{1}{(1 - t/\tau)^\gamma}.
    \label{eq:analytic1}
\end{equation}

\noindent Note that $\gamma$ can also be written as $\gamma = (m_3 - m_1) {bg}/({kT \Phi_1}) = {\phi_c}/({\Phi_1 x_1}) \approx \phi_c/\phi$.

\section{Analytic solution of D/H in a helium-dominated atmosphere} \label{sec:analytic2}

We present an analytic solution for D/H in a ternary mixture of H, D, and He to better understand the sharp transition to enhanced D/H in helium-dominated atmospheres shown in Figure \ref{fig:mean_DH}. For simplicity, we allow only D and H to escape from a static He background. The diffusion-limited escape flux for species $i$ in a He-dominated atmosphere is

\begin{equation}
    \Phi_\mathrm{i} = b x_i \left( \frac 1{H_{\mathrm{He}}} - \frac 1{H_i}\right)
\end{equation}

\noindent where $x_i = N_i / (N_{\mathrm{He}} + N_i)$ is molar concentration and $x_\mathrm{i} \approx N_i / N_\mathrm{He}$ for a minor species $i$. $b$ is the binary diffusion coefficient for He and species $i$. Combining with Equation \ref{eq:dNdt}, we get

\begin{equation}
    \frac{d \log N_i}{dt} \approx  - \frac{A b}{N_{He}} \left( \frac 1{H_{\mathrm{He}}} - \frac 1{H_i}\right).
\end{equation}

\noindent Expanding the scale height terms gives us

\begin{equation}
    \frac{d \log N_{\mathrm{H}}}{dt} \approx -\frac{A b}{N_\mathrm{He}} \left( \frac {4m_\mathrm{p} g}{kT} - \frac {m_\mathrm{p} g}{kT} \right)
\end{equation}

\noindent where $m_\mathrm{p}$ is the proton mass and $m_{\mathrm{He}} \approx 4m_\mathrm{p}$. This simplifies to

\begin{equation}
    \frac{d \log N_{\mathrm{H}}}{dt} \approx -\frac{3}{\tau_\mathrm{f}}
    \label{eq:dlogHdt}
\end{equation}

\noindent for H, where $\tau_\mathrm{f}$ is the fractionation timescale:

\begin{equation}
    \tau_\mathrm{f} \equiv \frac{N_\mathrm{He}}{A b}  \frac {kT}{m_\mathrm{p} g}. 
\end{equation}

\noindent The total number of moles of He in the atmosphere is

\begin{equation}
    N_{\mathrm{He}} \approx \frac{f_{ret}M_\mathrm{p}}{m_{\mathrm{He}}}.
\end{equation}

\noindent Substituting this into $\tau_\mathrm{f}$, we get

\begin{equation} \label{tau_f}
    \tau_\mathrm{f} = \frac{f_{ret} M_\mathrm{p}}{A b} \frac{kT}{4m_\mathrm{p}^2 g},
\end{equation}

\noindent which simplifies to:

\begin{equation}
    \tau_\mathrm{f} = \frac{f_{ret} k T}{16 \pi G m_\mathrm{p}^2 b}.
\end{equation}

\noindent Integrating Equation \ref{eq:dlogHdt} and a similar expression for D, $d \log N_\mathrm{D}/dt = -2/\tau_\mathrm{f}$, the number of moles of H and D at time $t$ is

\begin{equation}
    N_\mathrm{H} = N_\mathrm{H,0}^{ -3t/\tau_\mathrm{f}},\ N_\mathrm{D} = N_\mathrm{D,0}^{ -2t/\tau_\mathrm{f}}.
\end{equation}

\noindent To solve for D/H at time $t$, we simply divide these expressions:

\begin{equation} \label{tau_f2}
    \frac{N_\mathrm{D}}{N_\mathrm{H}} = \left[ \frac{N_\mathrm{D,0}}{N_\mathrm{H,0}} \right] ^{ t/\tau_\mathrm{f}}.
\end{equation}

\noindent The result is a classic expression for Rayleigh fractionation.

\end{document}